\documentclass[]{spie}

\usepackage{amsmath,amsfonts,amssymb}
\usepackage{graphicx}
\usepackage[colorlinks=true, allcolors=blue]{hyperref}

\def\bk{BICEP/\textit{Keck}}

\title{Development of the 220/270 GHz Receiver of BICEP Array}

\author[a,*]{Y.~Nakato}
\author[b]{P.~A.~R.~Ade}
\author[c]{Z.~Ahmed}
\author[d]{M.~Amiri}
\author[e]{D.~Barkats}
\author[f,h]{R.~Basu~Thakur}
\author[g]{C.~A.~Bischoff}
\author[a,c]{D.~Beck}
\author[f,h]{J.~J.~Bock}
\author[i]{V.~Buza}
\author[a]{B.~Cantrall}
\author[j]{J.~R.~Cheshire~IV}
\author[e]{J.~Cornelison}
\author[k]{M.~Crumrine}
\author[f]{A.~J.~Cukierman}
\author[l]{E.~Denison}
\author[e]{M.~Dierickx}
\author[m]{L.~Duband}
\author[e]{M.~Eiben}
\author[e,p]{B.~D.~Elwood}
\author[d,f]{S.~Fatigoni}
\author[n,o]{J.~P.~Filippini}
\author[a]{A.~Fortes}
\author[f]{M.~Gao}
\author[g]{C.~Giannakopoulos}
\author[a]{N.~Goeckner-Wald}
\author[c]{D.~C.~Goldfinger}
\author[a]{J.~A.~Grayson}
\author[e]{P.~K.~Grimes}
\author[k]{G.~Hall}
\author[a]{G.~Halal}
\author[d]{M.~Halpern}
\author[g]{E.~Hand}
\author[c]{S.~Henderson}
\author[l]{J.~Hubmayr}
\author[f]{H.~Hui}
\author[a,c,l]{K.~D.~Irwin}
\author[a,f]{J.~Kang}
\author[c]{K.~S.~Karkare}
\author[a]{E.~Karpel}
\author[f]{S.~Kefeli}
\author[e,p]{J.~M.~Kovac}
\author[a,c]{C.~L.~Kuo}
\author[f]{K.~Lau}
\author[g]{M.~Lautzenhiser}
\author[n]{A.~Lennox}
\author[a]{T.~Liu}
\author[h]{K.~G.~Megerian}
\author[e]{M.~Miller}
\author[f]{L.~Minutolo}
\author[f]{L.~Moncelsi}
\author[h]{H.~T.~Nguyen}
\author[f,h]{R.~O'Brient}
\author[f]{A.~Patel}
\author[e]{M.~A.~Petroff}
\author[e,p]{A.~R.~Polish}
\author[m]{T.~Prouve}
\author[k,j]{C.~Pryke}
\author[l]{C.~D.~Reintsema}
\author[f]{T.~Romand}
\author[a]{M.~Salatino}
\author[f]{A.~Schillaci}
\author[e,s]{B.~L.~Schmitt}
\author[j]{B.~Singari}
\author[f,h]{A.~Soliman}
\author[e,p]{T.~St.~Germaine}
\author[f]{A.~Steiger}
\author[f]{B.~Steinbach}
\author[b]{R.~Sudiwala}
\author[a,c]{K.~L.~Thompson}
\author[b]{C.~Tucker}
\author[h]{A.~D.~Turner}
\author[e]{C.~Verg\`{e}s}
\author[f]{A.~Wandui}
\author[h]{A.~C.~Weber}
\author[k]{J.~Willmert}
\author[c]{W.~L.~K.~Wu}
\author[a]{H.~Yang}
\author[a,c]{E.~Young}
\author[a]{C.~Yu}
\author[e]{L.~Zeng}
\author[a,f]{C.~Zhang}
\author[f]{S.~Zhang}

\affil[a]{Department of Physics, Stanford University, Stanford, CA 94305, USA}
\affil[b]{School of Physics and Astronomy, Cardiff University, Cardiff, CF24 3AA, United Kingdom}
\affil[c]{Kavli Institute for Particle Astrophysics and Cosmology, SLAC National Accelerator Laboratory, 2575 Sand Hill Rd, Menlo Park, CA 94025, USA}
\affil[d]{Department of Physics and Astronomy, University of British Columbia, Vancouver, British Columbia, V6T 1Z1, Canada}
\affil[e]{Center for Astrophysics, Harvard \& Smithsonian, Cambridge, MA 02138, USA}
\affil[f]{Department of Physics, California Institute of Technology, Pasadena, CA 91125, USA}
\affil[g]{Department of Physics, University of Cincinnati, Cincinnati, OH 45221, USA}
\affil[h]{Jet Propulsion Laboratory, Pasadena, CA 91109, USA}
\affil[i]{Kavli Institute for Cosmological Physics, University of Chicago, Chicago, IL 60637, USA}
\affil[j]{Minnesota Institute for Astrophysics, University of Minnesota, Minneapolis, MN 55455, USA}
\affil[k]{School of Physics and Astronomy, University of Minnesota, Minneapolis, MN 55455, USA}
\affil[l]{National Institute of Standards and Technology, Boulder, CO 80305, USA}
\affil[m]{Service des Basses Temp\'{e}ratures, Commissariat \`{a} l'Energie Atomique, 38054 Grenoble, France}
\affil[n]{Department of Physics, University of Illinois at Urbana-Champaign, Urbana, IL 61801, USA}
\affil[o]{Department of Astronomy, University of Illinois at Urbana-Champaign, Urbana, IL 61801, USA}
\affil[p]{Department of Physics, Harvard University, Cambridge, MA 02138, USA}
\affil[q]{Kavli Institute for the Physics and Mathematics of the Universe (WPI), UTIAS, The~University~of~Tokyo, Kashiwa, Chiba 277-8583, Japan}
\affil[r]{Aix-Marseille  Universit\'{e},  CNRS/IN2P3,  CPPM,  13288 Marseille,  France}
\affil[s]{Department of Physics and Astronomy, University of Pennsylvania, Philadelphia, PA 19104, USA}

\authorinfo{*Corresponding Author: Yuka Nakato\\E-mail: \href{mailto:yukanaka@stanford.edu}{yukanaka@stanford.edu}}

\begin{document} 
\maketitle

\begin{abstract}

Measurements of $B$-mode polarization in the CMB sourced from primordial gravitational waves would provide information on the energy scale of inflation and its potential form. To achieve these goals, one must carefully characterize the Galactic foregrounds, which can be distinguished from the CMB by conducting measurements at multiple frequencies. BICEP Array is the latest-generation multi-frequency instrument of the \bk\ program, which specifically targets degree-scale primordial $B$-modes in the CMB. In its final configuration, this telescope will consist of four small-aperture receivers, spanning frequency bands from 30 to 270 GHz. The 220/270 GHz receiver designed to characterize Galactic dust is currently undergoing commissioning at Stanford University and is scheduled to deploy to the South Pole during the 2024--2025 austral summer. Here, we will provide an overview of this high-frequency receiver and discuss the integration status and test results as it is being commissioned.
 
\end{abstract}

\keywords{Cosmic Microwave Background, Polarization, Instrumentation, Cosmology, Inflation}

\section{Introduction}
\label{sec:introduction} 

Precise measurements of the cosmic microwave background (CMB) play a key role in cosmology. In particular, we are motivated to measure the polarization patterns in the CMB to search for evidence of cosmic inflation. Under the $\Lambda$CDM standard model, assuming only scalar perturbations were present at the surface of last scattering, we expect the primordial CMB signal to consist purely of $E$-mode polarization patterns \cite{zaldarriaga_and_seljak}. However, inflationary theories additionally predict tensor perturbations in the early universe, i.e., a background of primordial gravitational waves. Tensor perturbations source both $E$-mode and $B$-mode patterns in the CMB via Thomson scattering at the recombination epoch \cite{seljak_and_zaldarriaga}. Thus by searching for an excess of $B$-modes in the CMB at recombination, we can detect or set limits on primordial gravitational waves from inflation. Since 2006, the \bk\ program has been taking measurements to constrain this primordial $B$-mode signature, parameterized by the tensor-to-scalar ratio, $r$. \bk’s latest published results, which includes \bk\ data taken up to the end of the 2018 season (``BK18''), set the most sensitive $B$-mode-only constraints to date on $r$, with a sensitivity of $\sigma(r) = 0.009$ \cite{bk18}.

The search for primordial gravitational waves is complicated by the presence of non-primordial $B$-mode signals in the CMB that come from Galactic foregrounds and gravitational lensing. We typically focus on two dominant types of polarized Galactic foreground contaminants: synchrotron radiation and thermal dust emission. These foreground components can be separated from the CMB through multi-frequency measurements utilizing their different spectra. While CMB signal peaks at around 160 GHz and follows a blackbody spectrum, Galactic polarized synchrotron emission follows a power law and is brightest at lower frequencies \cite{fuskeland}, and polarized thermal dust emission from within our Galaxy has a modified blackbody spectrum that dominates at higher frequencies \cite{bk_planck}.

In particular, polarized Galactic dust is brighter than synchrotron at the peak of the CMB spectrum in the region that BICEP is observing, and the sample variance from dust contributes more than 20\% of the total uncertainty in the best $B$-mode-only $r$ constraint to date \cite{bk18}. Thermal dust emission is caused by starlight re-radiating from dust grains in the interstellar medium. Because these dust grains are asymmetric and elongated, they align with the Galactic magnetic field and therefore emit slightly polarized photons. This emission is the brightest in the Galactic plane, highly anisotropic, and extends all the way to high Galactic latitudes where \bk\ is observing. In order to effectively guard for spurious $B$-modes from the Galaxy, any future experiment must carefully characterize the spectral properties of dust, which requires appropriate frequency coverage. Thus, a sensitive high-frequency instrument for taking precise measurements of the dust is a crucial part of the \bk\ program.

BICEP Array (BA) is the latest-generation multi-frequency instrument of the \bk\ program, replacing \textit{Keck Array}. BA, which specifically targets degree-scale $B$-modes in the CMB, will consist of four small-aperture receivers in its final configuration, spanning frequency bands from 30 to 270 GHz and covering the same sky patch as BICEP3. Two of these have already been deployed: a 30/40 GHz receiver for constraining synchrotron and a 150 GHz receiver observing close to the peak of the CMB spectrum. The other two slots of the telescope are currently occupied by \textit{Keck Array}\ receivers until BA receivers are ready to replace them. The next receiver of BA to deploy will be the 220/270 GHz receiver, which is designed to characterize Galactic dust. With a bigger field of view and larger detector count compared to the \textit{Keck}\ 220 GHz receiver that it will be replacing, this receiver will greatly improve the signal-to-noise of our dust measurement. This high-frequency receiver is currently undergoing commissioning at Stanford University and is scheduled to deploy to the South Pole during the 2024--2025 austral summer.

In this paper, we will focus on the 220/270 GHz receiver and first present an overview for this high-frequency instrument in Sec.~\ref{sec:ba4}. We then describe in detail in Sec.~\ref{sec:performance} the receiver integration status and performance as it is being commissioned for deployment in North America. We conclude by outlining the predicted overall impact and performance of this receiver on-site at the South Pole in Sec.~\ref{sec:conclusion}.

\section{220/270 GHz Instrument Overview}
\label{sec:ba4}

\begin{figure} [b]
\centering
\includegraphics[width=1.0\textwidth]{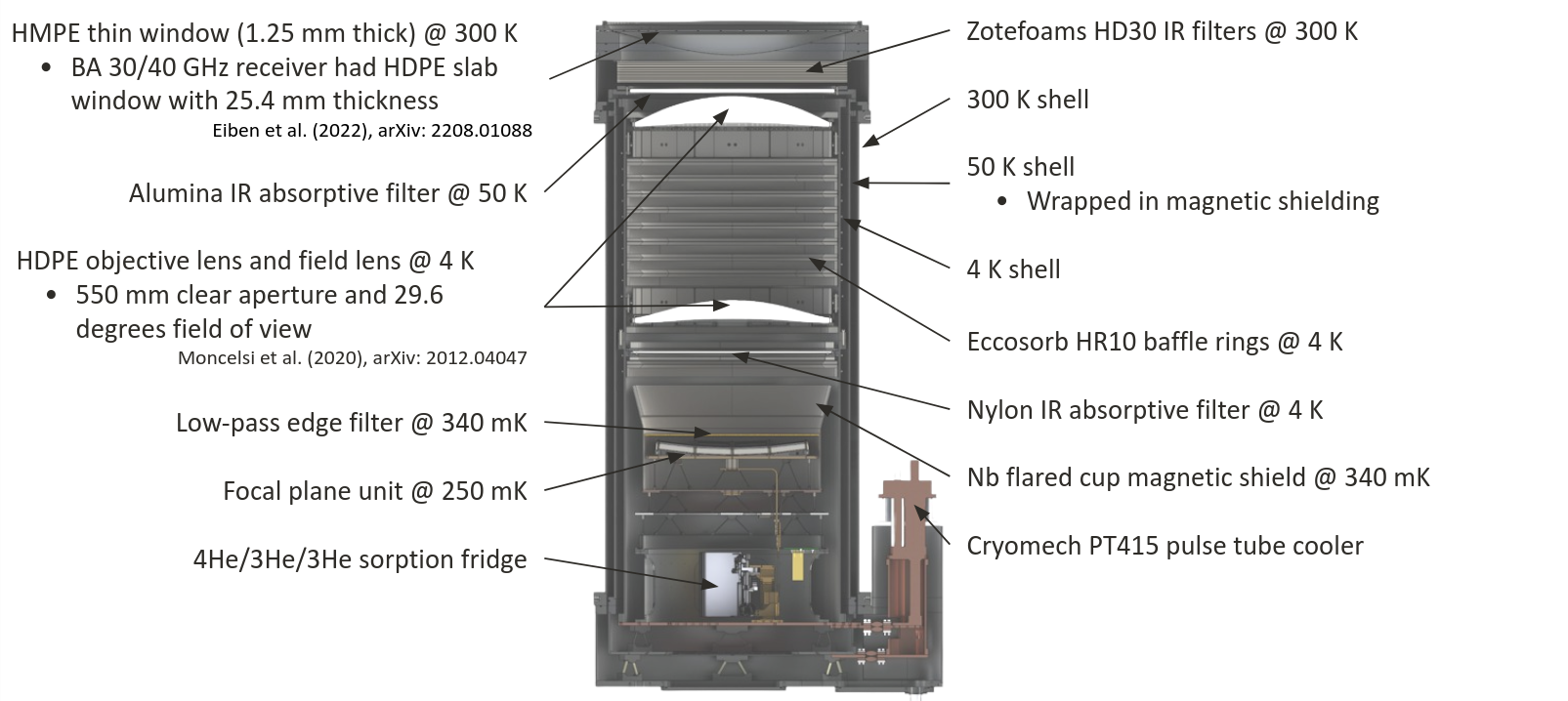}
\caption
{\label{fig:ba4_schematic_labeled} 
Schematic of the 220/270 GHz receiver.}
\end{figure}

Fig.~\ref{fig:ba4_schematic_labeled} shows a cross section of the 220/270 GHz receiver.

\subsection{Cryostat}
\label{sec:cryostat}

The 220/270 GHz receiver, like the other BA receivers, consists of an outer 300 K vacuum shell containing two nested stages with nominal operating temperatures of 50 K and 4 K \cite{mike_spie}. The stages are cooled using a Cryomech PT415 pulse tube, which has a cooling capacity of 40 W at 45 K and 1.5 W at 4.2 K.

Inside the 4 K shell, we have a three-stage 4He/3He/3He sorption fridge from CEA Grenoble, providing the cooling power for the three sub-Kelvin stages at 2 K, 340 mK, and 250 mK, respectively. This fridge is designed to achieve a hold time of over 48 hours for 15 $\mu$W load on the 250 mK stage (on-sky, we expect to have less than 8 $\mu$W total load on the focal plane). The sub-Kelvin stages are stacked ``wedding cake''-style, with each stage separated by low-conductivity carbon fiber truss legs, and the focal plane at the top of the structure containing the detector modules.

\subsection{Optics}
\label{sec:optics}

At the top of the 300 K vacuum shell, we have the 300 K laminate high modulus polyethylene (HMPE) thin vacuum window, which is 1.25 mm thick \cite{thin_window}. For the high frequencies that this receiver will observe at, using a thin window allows us to significantly reduce in-band emissions, and therefore instrumental photon noise on the detectors. HMPE laminates have been measured to be 100 times stronger than bulk high density polyethylene (HDPE), allowing them to get thin enough to act as an additional anti-reflection layer, while still being able to hold vacuum reliably \cite{denis_spie, thin_window}. We are using layered expanded polytetrafluoroethylene (ePTFE) for the anti-reflection layers, which in combination with the 1.25 mm thick window should result in a band average reflection less than 0.5\% \cite{miranda_spie}.

Below the vacuum window, there are various infrared filters in place to reduce radiative heat load onto the colder stages. At the very top, right under the window, is a stack of Zotefoams HD30 filters; underneath that at 50 K is an alumina infrared absorptive filter that is anti-reflection coated with two layers of epoxy on each side. At the 4K stage we have a Nylon infrared absorptive filter, and at the sub-K stage is a low-pass metal mesh filter \cite{lpe}.

The lenses are made of HDPE and cooled to 4 K to minimize loading on the detectors. The objective lens is located at the top of the 4 K shell, and the field lens is in the middle. Eccosorb HR10 baffle rings populate the area between the lenses, for the purpose of minimizing in-band reflections.

\subsection{Detector Modules and Readout}
\label{sec:detectors}

The 220/270 GHz BA receiver is a dual-band receiver with a ``checkerboard'' of 220 and 270 GHz single-band detector modules on its focal plane. The focal plane is ``curved'' for this receiver, identical in configuration to the 150 GHz BA receiver. Here, the modules themselves are flat as in all the previous BA receivers \cite{lorenzo_spie}, but they are mounted using spacers to approximate a 1.5 m radius of curvature spherical focal plane surface for improved optical performance. We have 12 modules for a full focal plane, and each module has a 6 inch silicon wafer with 18 by 18 pixels, with 2 detectors per pixel for the two orthogonal polarizations, which gives a total of 7776 detectors, matching the high detector count of the 150 GHz receiver. This is a significant upgrade from the \textit{Keck}\ 220 GHz receiver that this receiver will be replacing, which has only 512 detectors.

\begin{figure} [b]
\begin{center}
\begin{tabular}{c} 
\includegraphics[width=0.65\textwidth]{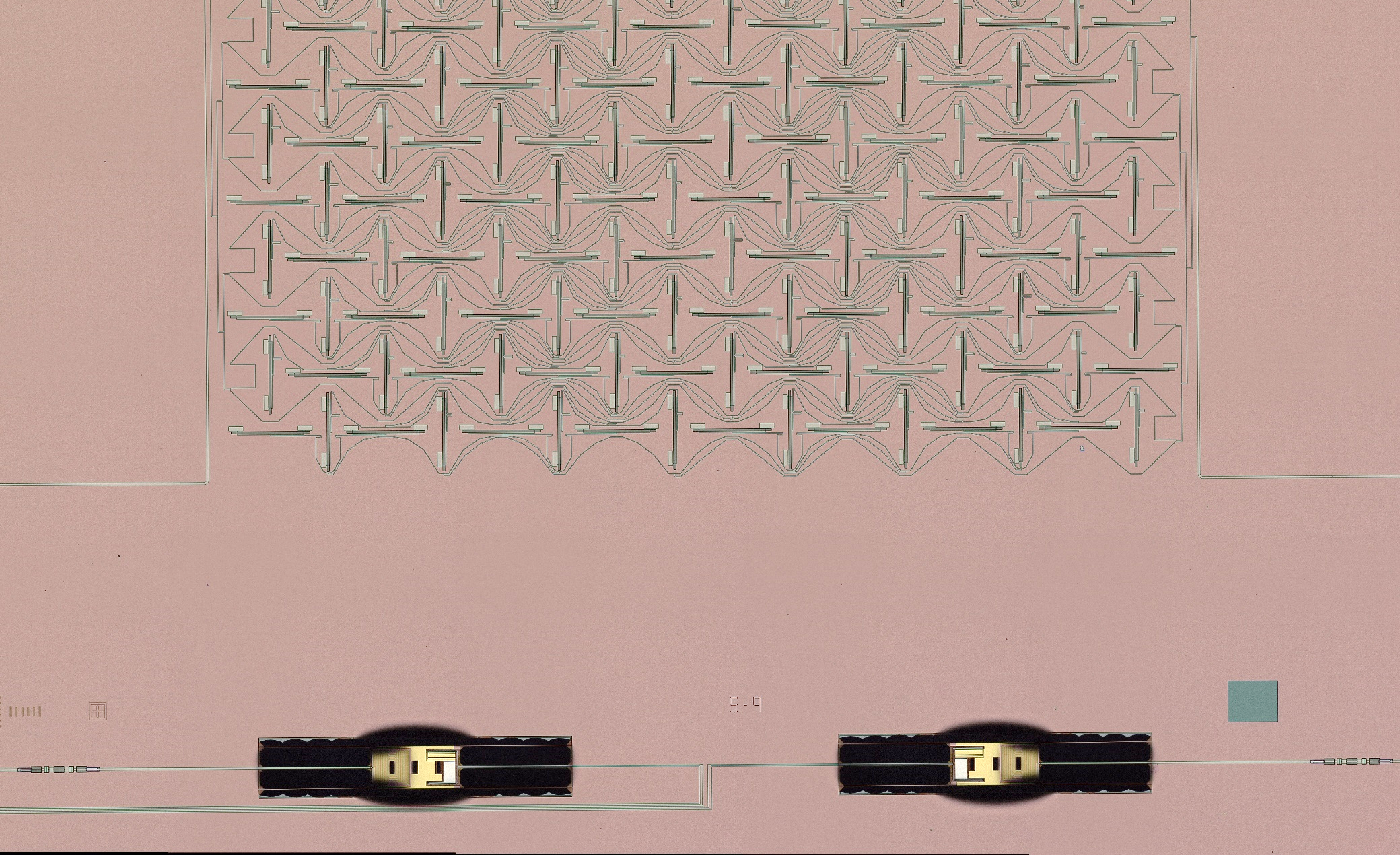}
\end{tabular}
\end{center}
\caption
{\label{fig:H9_TES_cropped} 
Partial view of a single 220 GHz pixel for the 220/270 GHz receiver.}
\end{figure}

For our detectors, we use dual-polarization antenna-coupled TES arrays. Fig.~\ref{fig:H9_TES_cropped} shows a partial view of a single 220 GHz pixel for the high-frequency BA receiver. At the top is part of the antenna network. The signal in the vertical slots are fed into the ``A'' detector, and similarly the signal in the horizontal slots goes to the ``B'' detector. In this way, we are able to measure orthogonal ``A'' and ``B'' polarizations of the same location in the sky. Below the antenna network are two TES islands, each paired with a bandpass filter, as seen on the left and right edges at the bottom of the figure. The signal goes through the bandpass filter before getting transmitted to the TES island. It is this bandpass filter that defines the upper and lower frequency cutoffs of the science band.

We read out the detectors using time-division multiplexing multi-channel electronics (MCE) readout developed by the University of British Columbia \cite{sofia_mce}. We have a multiplexing factor of 41 for this receiver, matching the 150 GHz receiver, read out through two stages of SQUIDs made by NIST.

\section{Performance}
\label{sec:performance}

The commissioning of the 220/270 GHz cryostat at Stanford University began in 2021. Between mid-2021 and early 2023, multiple cold tests were performed to check the vacuum, thermometry readout, and the cryogenic performance of the receiver as more components were added to it. Starting out with an ``empty'' receiver, the 4He/3He/3He sorption fridge was the first to be added and checked, then the sub-Kelvin stages, all cabling, and finally the optics and detector modules. We achieved good cryogenic performance with all components installed in 2023. With all optics installed and the receiver open to light, the sorption fridge hold time is well over 48 hours before needing to be cycled, and we achieve a focal plane base temperature of 270 mK. For the rest of 2023 and all of 2024, this receiver has been undergoing various tests of the detectors, optics, and readout.

\subsection{Integrated Receiver Testing with the First 220 GHz Modules}
\label{sec:detector_testing}

The first optical test runs in the 220/270 GHz receiver were conducted in 2024. In these runs, we tested the first two 220 GHz detector modules fabricated at Caltech/JPL, which we call ``H10'' and ``H9'', in the receiver outfitted with all official optics. Our optical tests include measurements of end-to-end optical efficiency, spectral response, and near-field beam patterns.

\subsubsection{End-to-End Optical Efficiency}
\label{sec:oe}

\begin{figure} [b]
\begin{center}
\begin{tabular}{c} 
\includegraphics[width=0.75\textwidth]{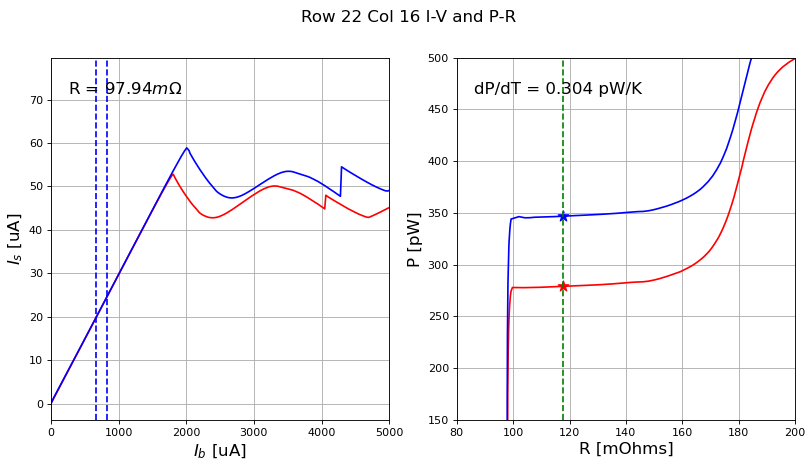}
\end{tabular}
\end{center}
\caption
{\label{fig:load_curve} 
A typical Al load curve for detectors on the 220 GHz tile H10. Load curves are taken by first driving the TESs normal by applying a high bias voltage, then stepping down in bias until the detectors are superconducting again.
Red lines show the 300 K measurements, while blue lines show the 77 K measurements.
Left: Current through the TES as a function of the bias current. As we go down in bias (right to left), the Al TES can be seen to be transitioning from the normal state to the superconducting state.
Right: The electrical power as a function of TES resistance.}
\end{figure}

We measure the end-to-end optical efficiency (OE) for the two 220 GHz detector modules H10 and H9 in the 220/270 GHz cryostat. This test is done by looking at the detector response for two different beam-filling optical loads, at ambient temperature (300 K) and liquid nitrogen temperature (77 K). For each of the two optical loads, we take load curves, where we measure the TES current while ramping bias current across the Al transition. A typical load curve for a single TES is shown in Fig.~\ref{fig:load_curve}.

\begin{figure} [t]
\begin{center}
\begin{tabular}{c} 
\includegraphics[width=0.95\textwidth]{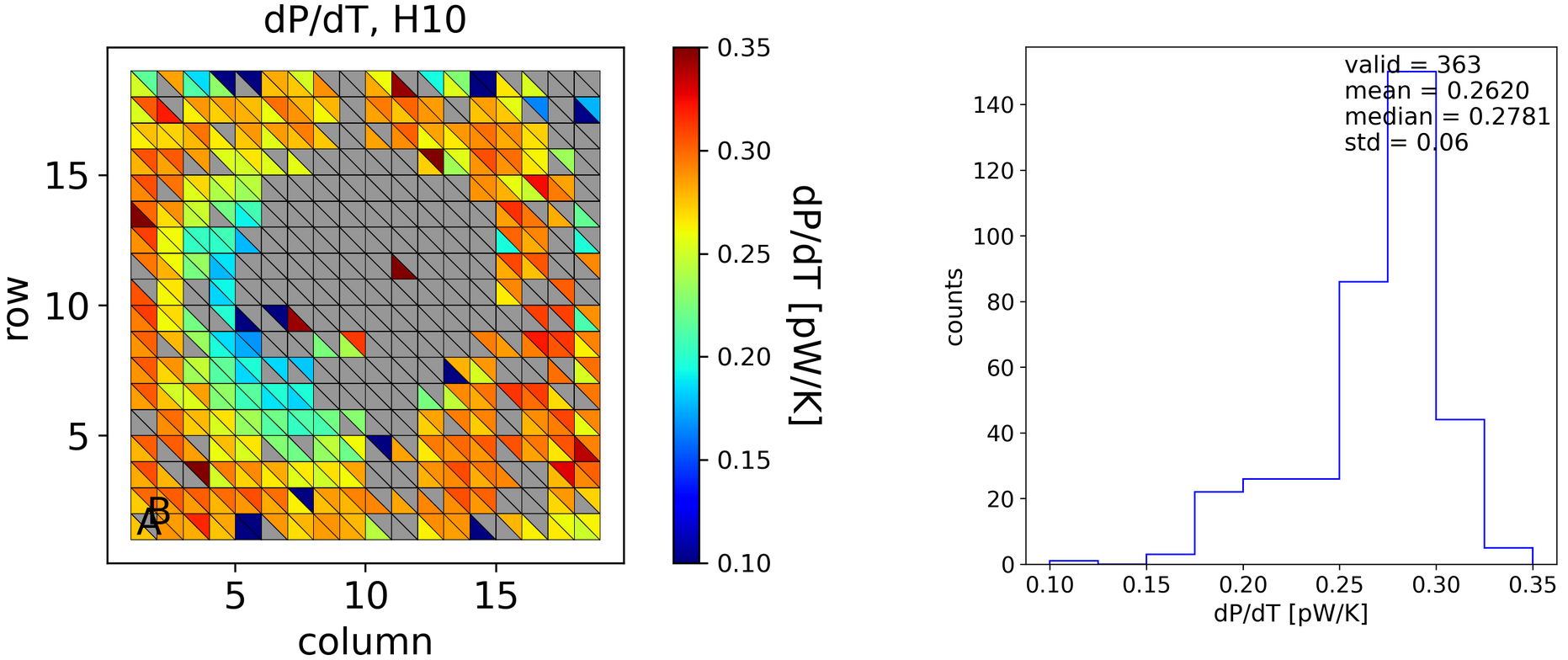}
\end{tabular}
\end{center}
\caption
{\label{fig:h10_dpdt} 
Tile map and histogram of the $\mathrm{d}P_{\mathrm{opt}}/\mathrm{d}T$ results for the first BA 220 GHz tile, H10. The center of H10 has dead detectors that cannot go into transition, due to a one-time fabrication error, where the TES islands were shorted to the heat sink, therefore making the TESs always cold and superconducting. The histogram excludes the dead detectors in the center.
}
\end{figure}

\begin{figure} [t]
\begin{center}
\begin{tabular}{c} 
\includegraphics[width=0.95\textwidth]{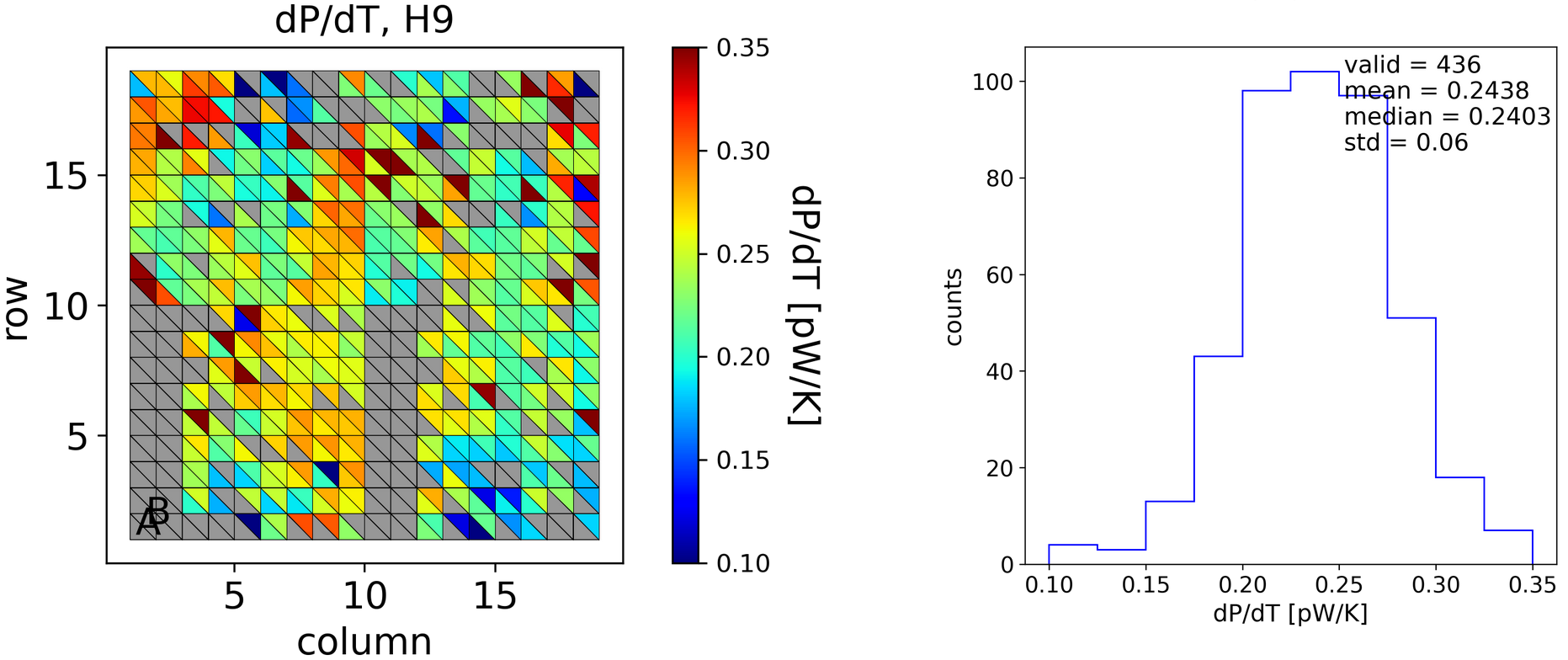}
\end{tabular}
\end{center}
\caption
{\label{fig:h9_dpdt} 
Tile map and histogram of the $\mathrm{d}P_{\mathrm{opt}}/\mathrm{d}T$ results for the second BA 220 GHz tile, H9. The dead half-columns are due to SQUID problems, which will be fixed before deployment by replacing the corresponding SQUID chips.
}
\end{figure}

From the load curves, we extract the $\mathrm{d}P_{\mathrm{opt}}/\mathrm{d}T$, the change in optical power deposited on the detectors with respect to optical load temperature. The resulting $\mathrm{d}P_{\mathrm{opt}}/\mathrm{d}T$ values for every detector of the two tiles are shown in the tile maps in Fig.~\ref{fig:h10_dpdt} and Fig.~\ref{fig:h9_dpdt}. Here, it is evident that the center of H10 has a ``hole'' of dead TESs that cannot go into transition. This is due to a one-time fabrication error, and the error was fixed for H9 and future wafers. We plan to replace the tile in H10 before deployment.

From the histograms, we obtain a median $\mathrm{d}P_{\mathrm{opt}}/\mathrm{d}T$ of around 0.28 pW/K for H10, and 0.24 pW/K for H9. Assuming the Rayleigh--Jeans limit, we use $\mathrm{d}P_{\mathrm{opt}}/\mathrm{d}T = k_B \eta \Delta\nu$ where $k_B$ is Boltzmann's constant, $\eta$ is the percent efficiency, and $\Delta\nu$ is the bandwidth. With $\Delta\nu = 63$ GHz for H10 and $\Delta\nu = 60$ GHz for H9 as obtained from our FTS measurements (see Sec.~\ref{sec:fts}), this gives us a percent efficiency of $\eta = 32\%$ and $\eta = 29\%$ for H10 and H9, respectively. This meets our design target of around 30\% optical efficiency, including losses from all optics.

\subsubsection{Fourier Transform Spectroscopy}
\label{sec:fts}

\begin{figure} [t]
\begin{center}
\begin{tabular}{c} 
\includegraphics[width=0.45\textwidth]{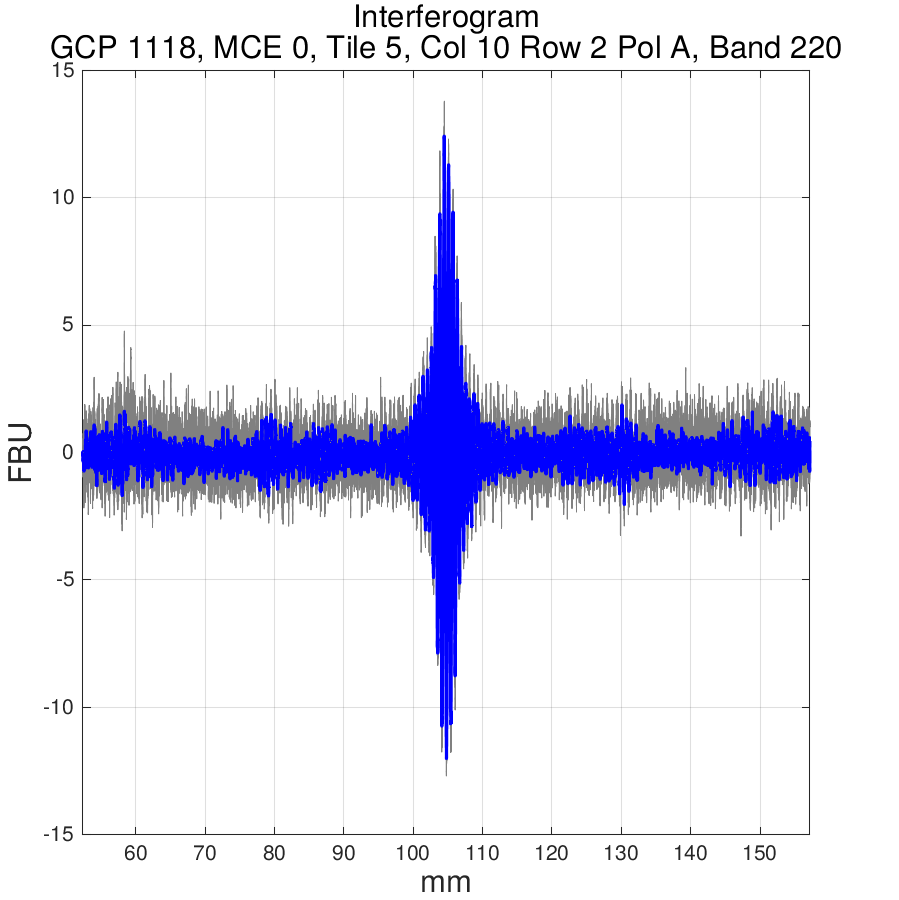}
\end{tabular}
\end{center}
\caption
{\label{fig:interferogram} 
Typical interferogram in the FTS test for the first BA 220 GHz tile, H10. The grey lines show each of the 6 scans of the dataset, and blue shows the median.}
\end{figure}

Next, we measure the spectral response of the 220 GHz detector modules with Fourier transform spectroscopy (FTS), using a custom-built Martin--Puplett interferometer that we mount directly to the receiver window. Fig.~\ref{fig:interferogram} shows an interferogram of a typical detector on these modules.

\begin{figure} [b]
\begin{center}
\begin{tabular}{c} 
\includegraphics[width=0.85\textwidth]{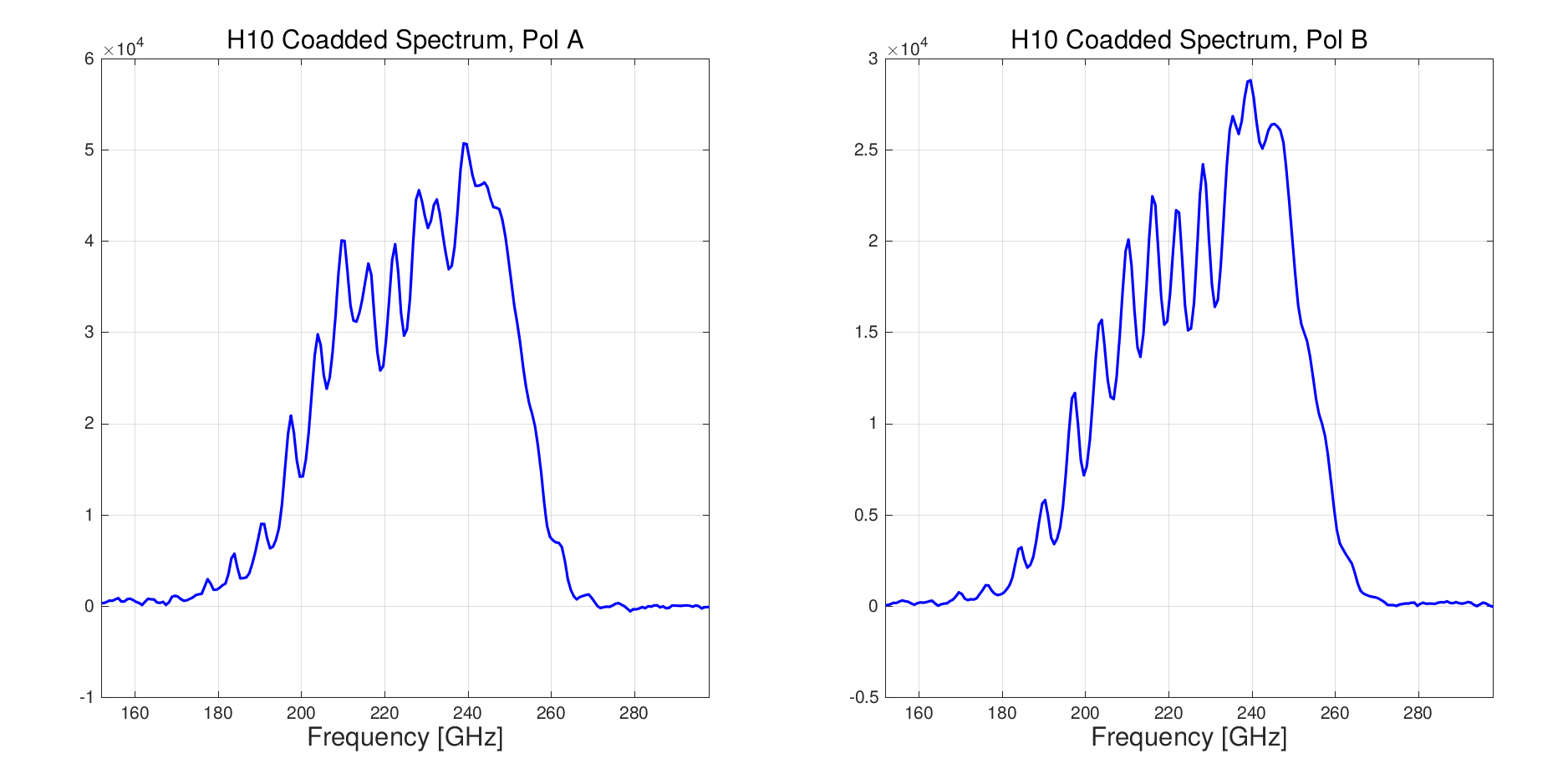}
\end{tabular}
\end{center}
\caption
{\label{fig:H10_spec} 
H10 co-added spectral response for polarization A (left) and polarization B (right).}
\end{figure}

\begin{figure} [t]
\begin{center}
\begin{tabular}{c} 
\includegraphics[width=0.85\textwidth]{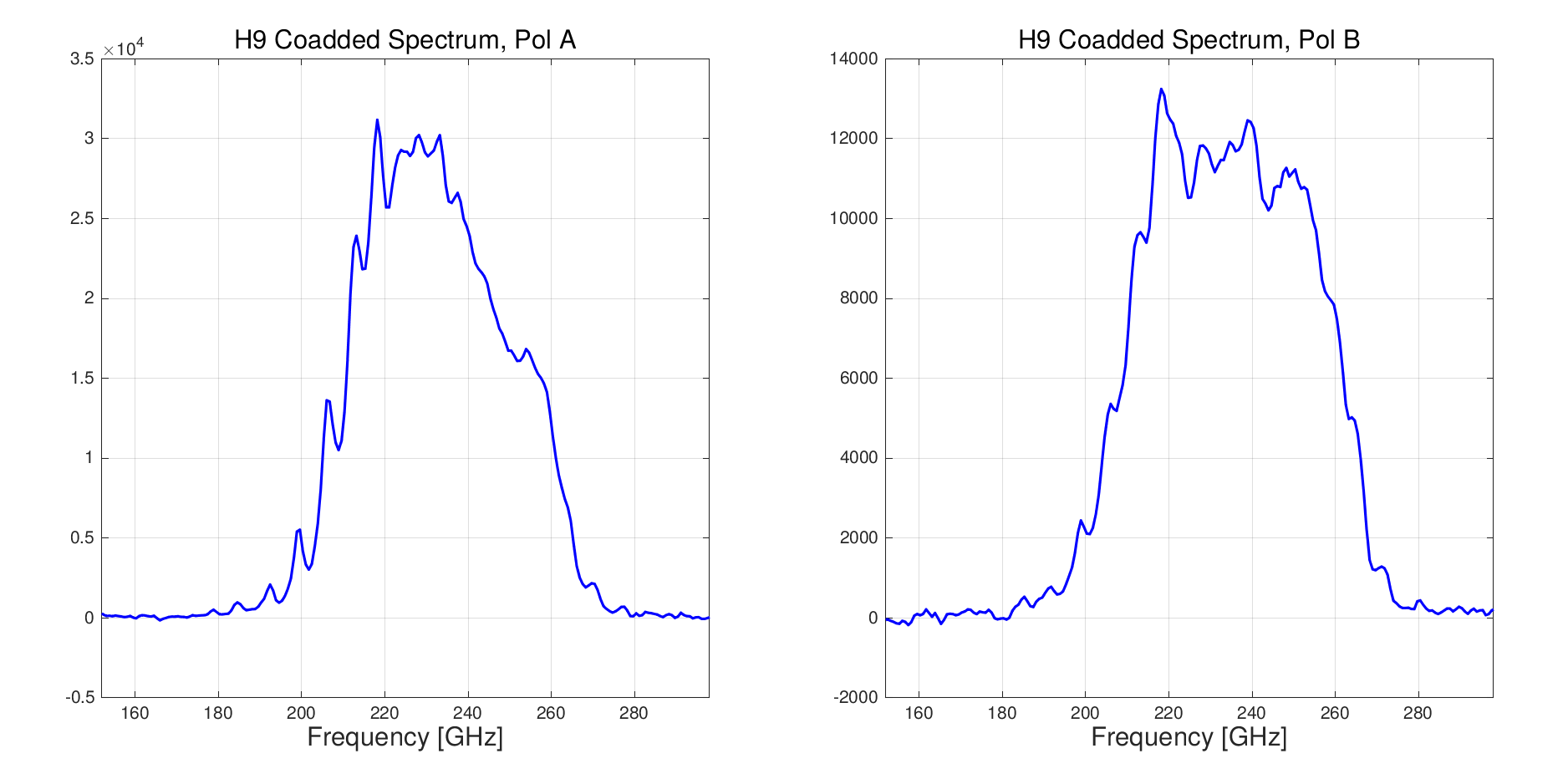}
\end{tabular}
\end{center}
\caption
{\label{fig:H9_spec} 
H9 co-added spectral response for polarization A (left) and polarization B (right).}
\end{figure}

We Fourier transform the interferograms, and the plots for the resulting co-added spectral response for each module are shown in Fig.~\ref{fig:H10_spec} and Fig.~\ref{fig:H9_spec}. The fringing in the H10 spectral response that shows up primarily on the low frequency side of the band is caused by a design error on the pixels of this detector tile. Narrower than intended microstrips were erroneously added to either side of the bandpass filters on the pixels, causing degraded transmission. This error has since been fixed for all future detector tiles including H9.

For H10, we measure a median band center and bandwidth of 228 GHz and 63 GHz (28\%), respectively.
For H9, we measure 233 GHz for the band center and 60 GHz (26\%) for the bandwidth.
We have also checked that this result is uniform across the entire tile.

\subsubsection{Near Field Beam Mapping}
\label{sec:nfbm}

As a final check of the overall health and performance of our optics, we take some beam maps with a source in the near field. For this near field beam mapping (NFBM) test we use a chopped thermal source mounted above the receiver window, which is moved around with a pair of Velmex stepper motors to make a map of the detector response. 

\begin{figure} [b]
\begin{center}
\begin{tabular}{c} 
\includegraphics[width=0.75\textwidth]{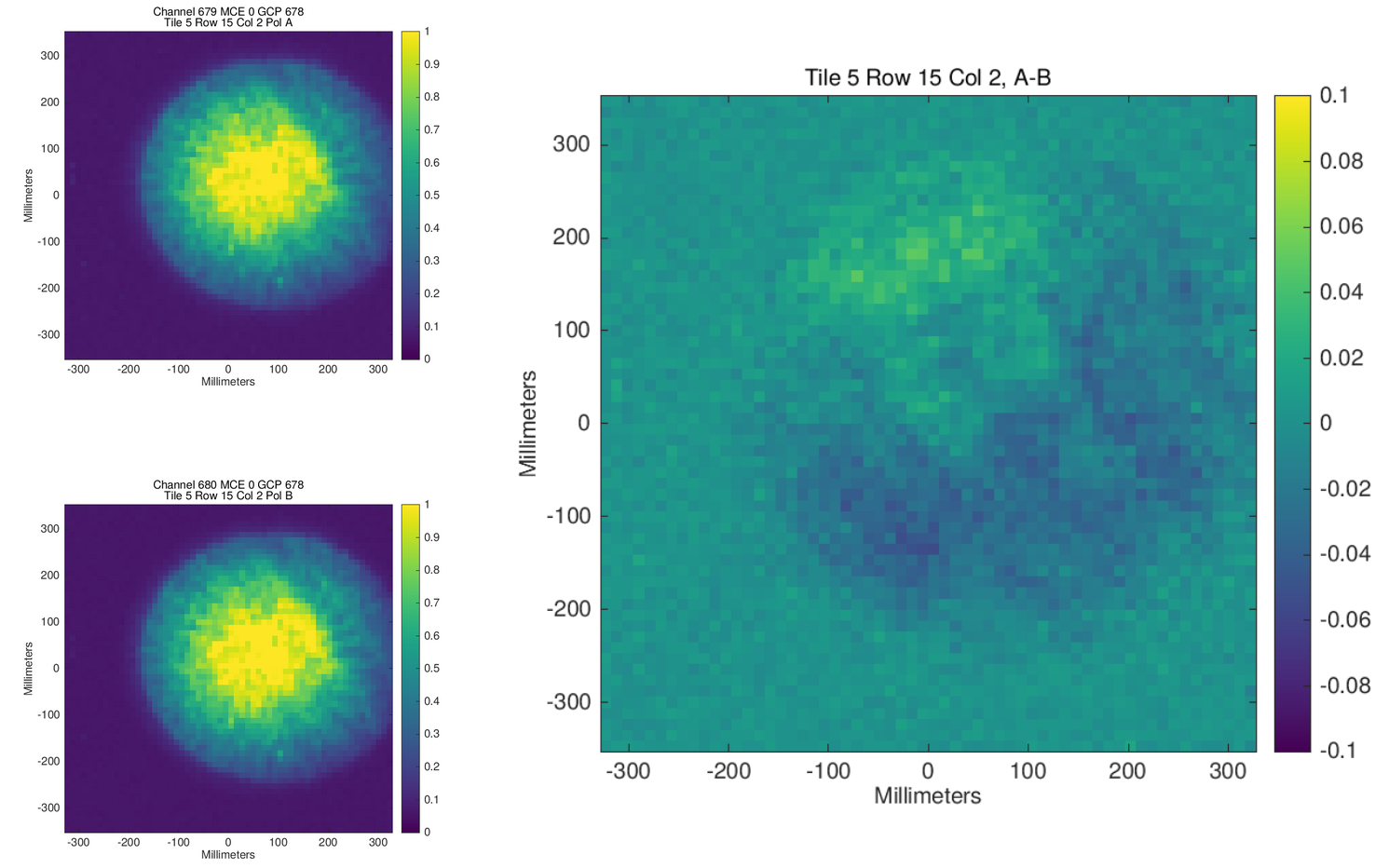}
\end{tabular}
\end{center}
\caption
{\label{fig:beam_map} 
Typical beam maps of a detector pair close to the edge of module H10. Left: Beam maps for a typical detector pair, showing polarization A (top) and polarization B (bottom). Right: The beam map difference between the two polarizations.}
\end{figure}

Fig.~\ref{fig:beam_map} shows beam maps for a typical detector pair. The beams are concentric with good signal-to-noise, and the beam shapes for the two polarizations match.

\section{Conclusion}
\label{sec:conclusion}

In-lab optical testing of the 220/270 GHz receiver was performed at Stanford University. This testing involved the first two 220 GHz detector modules, the complete set of official optics, and all necessary cabling. The goals for this test run was to characterize the optics and the detector modules, assess the cryogenic performance, and verify the readout. These tests in North America provide insight of the predicted overall performance of the receiver on-site at the South Pole.

\begin{figure} [b]
\begin{center}
\begin{tabular}{c} 
\includegraphics[width=0.9\textwidth]{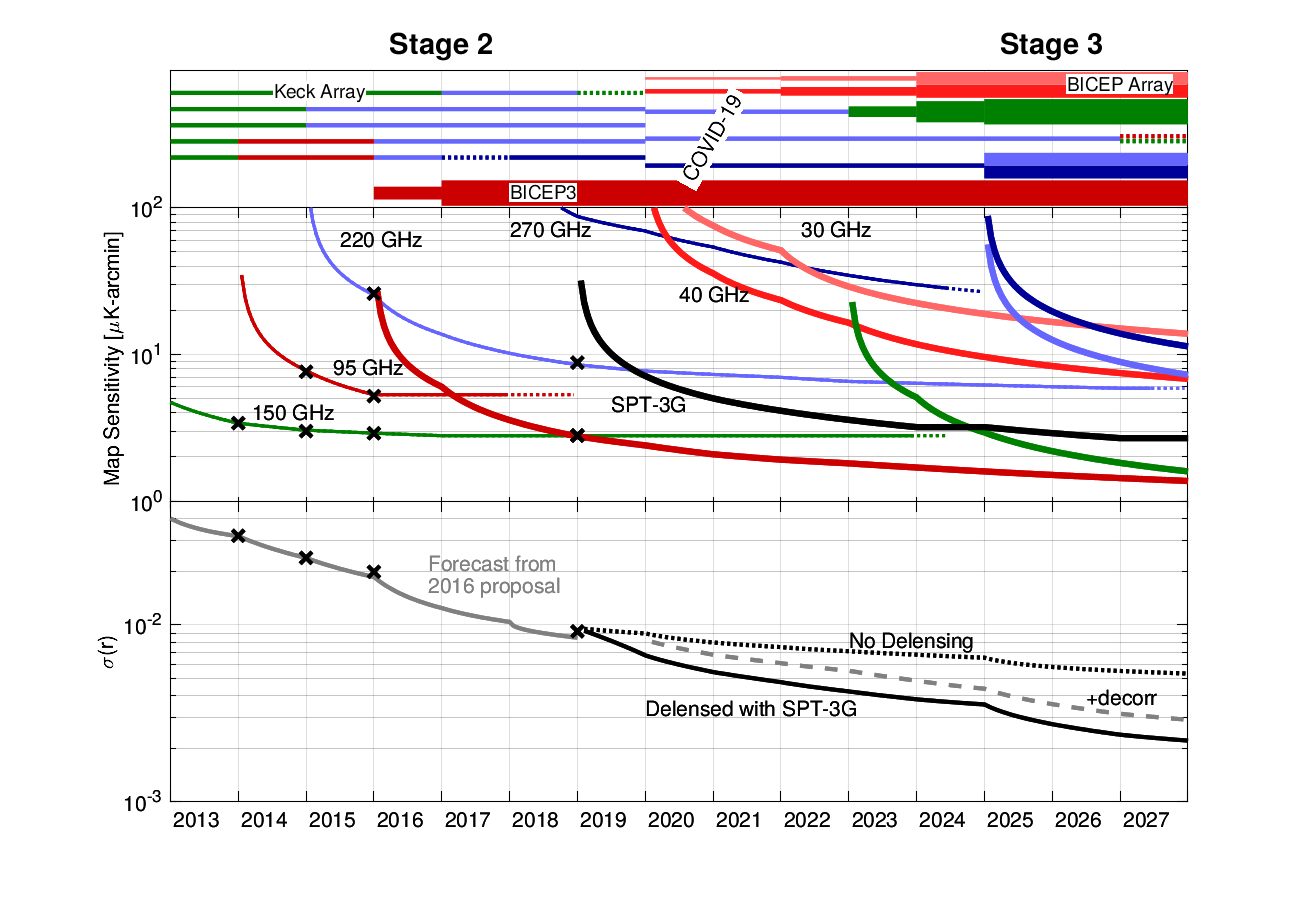}
\end{tabular}
\end{center}
\caption
{\label{fig:proj} 
Achieved and projected sensitivity of the \bk\ program, adapted from Fig.~5 of Cheshire et al. (2024) \cite{jamie_moriond}.
Top: The frequency bands covered by the various receivers through each season, with the different colors showing the different frequencies, and the width representing detector count.
Middle: Achieved and projected map sensitivities for each frequency band. SPT-3G is included in the solid black line. Black $\times$ symbols indicate published values.
Bottom: Progression of our sensitivity to $r$ over time. The solid line shows the projected $\sigma(r)$ with delensing in collaboration with SPT-3G, and the dotted line is the case with no delensing.
}
\end{figure}

The 220/270 GHz receiver has excellent cryogenic performance, and the readout has been tested. The results of the OE, FTS, and NFBM tests are all within spec, and we are on track to deploy to the South Pole in the 2024--2025 austral summer season. Fig.~\ref{fig:proj} is a schematic of the achieved and projected sensitivity for the entire \bk\ program, and shows the predicted impact of this high frequency receiver as it starts observing at the end of 2024. In the top subplot, the 220/270 GHz BA receiver can be seen starting operations in blue, in the 2025 observing season. Until this, the high frequency band is only covered by \textit{Keck}\ receivers.

With data up to and including 2027, without delensing, we estimate that our uncertainty on $r$ will be reduced to $\sigma(r) \sim 0.006$ \cite{bk_sptpol}. With delensing, we expect to achieve $\sigma(r) \sim 0.003$ \cite{bk18}. Even without delensing, it is clear that there is a ``cusp'' of improved $\sigma(r)$ in 2025 with the 220/270 GHz receiver added as seen in the bottom subplot of Fig.~\ref{fig:proj}, as this receiver will significantly improve our sensitivity to Galactic dust.

More high frequency detector modules are currently being fabricated at Caltech/JPL, and we will run more optical tests in North America before deployment to characterize these modules.

\appendix

\acknowledgments

The \bk\ projects have been made possible through a series of grants from
the U.S. National Science Foundation and by the Keck Foundation. The development of
antenna-coupled detector technology was supported by the JPL Research and Technology Development Fund and NASA Grants. Readout electronics were supported by a Canada Foundation for Innovation grant to UBC. The analysis effort at Stanford and SLAC is partially supported by the U.S. DOE Office of Science.
The computations in this paper were run on the Cannon cluster supported by the
FAS Science Division Research Computing Group at Harvard University. We thank
the staff of the U.S. Antarctic Program and in particular the South Pole Station without whose help this research would not have been possible. Special thanks go to our heroic winter-overs.

\bibliography{bibliography}
\bibliographystyle{spiebib} 

\end{document}